\documentclass[prl,aps,twocolumn,showpacs,groupedaddress,superscriptaddress]{revtex4}

\usepackage{graphicx,epsf,bm,bbm,color,amsmath,ulem,amssymb}

  \newcommand{\tr}[1]{\textrm{#1}}

	\newcommand{\beq}{\begin{eqnarray}}
	
	\newcommand{\eeq}{\end{eqnarray}}
	\newcommand{\nn}{\nonumber}

\newcommand{\bem}{\begin{pmatrix}}
\newcommand{\eem}{\end{pmatrix}}
\newcommand{\hb}{\hbar}

\newcommand{\f}{\frac}
\newcommand{\mbf}[1]{\mathbf{#1}}

\begin{document}
	
		\title{Pseudospin Vortex Ring with a Nodal Line in Three Dimensions}
        \author{Lih-King Lim}
        \affiliation{Institute for Advanced Study, Tsinghua University, Beijing 100084, People's Republic of China}
				\affiliation{Max-Planck-Institut f\"ur Physik komplexer Systeme, D-01187 Dresden, Germany}
        \author{Roderich Moessner}
        \affiliation{Max-Planck-Institut f\"ur Physik komplexer Systeme, D-01187 Dresden, Germany}
\begin{abstract}
We present a model of a topological semimetal in three dimensions whose energy spectrum exhibits a nodal line acting as a vortex ring; this in turn is linked by a pseudospin structure akin to that of a smoke ring. Contrary to a Weyl point node spectrum, the  vortex ring gives rise to skyrmionic pseudospin patterns in cuts on both sides of the nodal ring plane; this pattern covers the full Brillouin zone, thus leading to a fully extended chiral Fermi arc and a new, `maximal', anomalous Hall effect in a 3D semimetal. Tuning a model parameter shrinks the vortex ring until it vanishes, giving way to a pair of Weyl nodes of  opposite chirality.  This establishes a connection between two distinct momentum-space topologies - that of a vortex ring (a circle of singularity) and a monopole-anti-monopole pair (two point singularities). We present the model both as a low-energy continuum and a two-band tight-binding lattice model. Its simplicity permits an analytical computation of its Landau level spectrum.
\end{abstract}
	\maketitle

{\it Introduction\,-- }
The fruitful search for topological materials now extends beyond insulators. A most prominent example is the Weyl semimetal, which despite the gaplessness of  its bulk, hosts topologically protected surface states \cite{Murakami07,Wan11,Weng15a,Huang15,Xu15,Lu15,Lv15a,Lv15b,Yang15}. The central conceptual shift  is  from the energy band dispersion to the singularity structure in  momentum space. In this spirit, in graphene, the two-dimensional (2D) Dirac fermion originates from the pseudospin vortex texture, giving rise to the famous $\pi$ Berry phase physics \cite{Novo05,Gusynin05,Zhang05}. Its generalisation is the three-dimensional (3D) Weyl fermion, which emanates from a pseudospin monopole \cite{Volovik87,Wan11,Volovik03}. The latter acts as the termination of the topological Fermi arc \cite{Wan11} and gives rise to an intrinsic, albeit unquantized, anomalous Hall effect (AHE), a condensed matter phenomenon unique in 3D Weyl semimetals \cite{Haldane04,Klimkhamer05,Yang11,Burkov11a}.

The key diagnostic of topological semimetals remains the familiar one borrowed from band topology for a 2D Chern insulator, namely the Chern number reflected in the physical Hall response \cite{Thouless82,Hasan10,Qi11}. Continuing with the Weyl fermion example, when confining a pseudospin monopole in the 3D Brillouin zone, one is led to a planar Chern number that changes discontinuously from 0 to 1 as the point singularity is crossed \cite{Wan11}. In other words, the embedding of the point singularity in  3D momentum space leads to stacks of 2D skyrmionic pseudospin textures \cite{Volovik03} on only one, but not the other, side of the singularity. This we call a planar Chern composition (PCC) rule corresponding to the pseudospin monopole.

\begin{figure}
\begin{center}
\includegraphics[width=8.2cm]{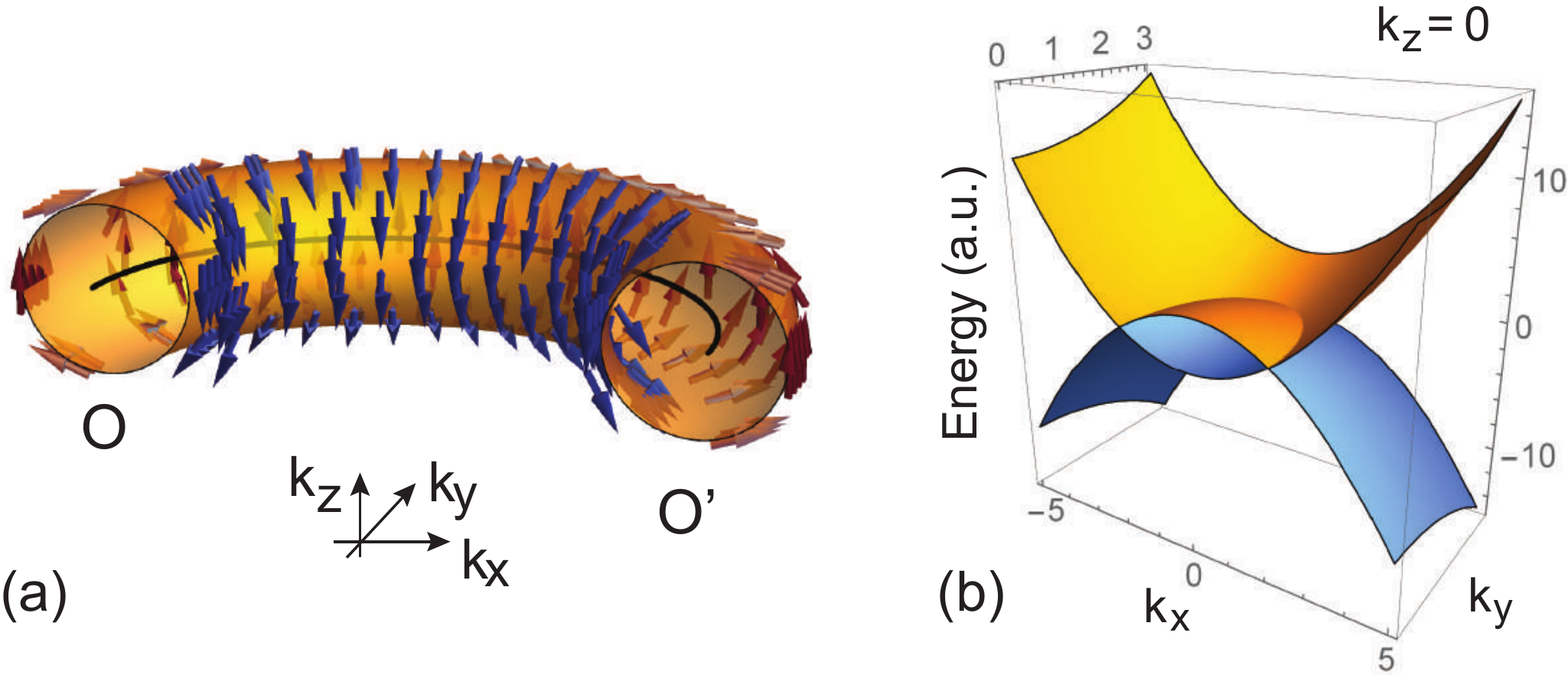}
\end{center}
\caption{(a) Smoke ring pseudospin structure shown on the toroidal Fermi surface close to the nodal ring.  (b) The low-energy nodal ring energy spectrum for $k_z=0$. For clarity, both panels are shown with a plane cut through the origin. }
\end{figure}

Here we construct a new band structure that shows that the pseudospin monopole PCC is not a unique one. Our analysis is motivated by recent interest in a new class of symmetry-protected nodal ring semimetals \cite{Burkov11b,Phillips14,Weng15b,Fang15,Mullen15,Kim15,Yu15,Chen15,Rhim15,Heikkila15,Xie15,Chan15,Bian15,Bian16,Yamakage16,Ezawa16,Wang16,Bzdusek16,Yan16,Chan16}, where a line node that forms a closed loop in the energy spectrum. Similarly to the Dirac points in graphene \cite{Hasan10}, this feature requires  symmetry protection.

This circular loop energy degeneracy opens the door to an extended, rather than point-like, singularity structure, which we construct as follows. First, inspired by the smoke ring in vortex dynamics \cite{Saffman92,Cooper99} and studies of 2D graphene-bilayer with higher winding vortices \cite{McCann06, Gail11}, we directly construct a class of pseudospin Hamiltonians exhibiting a vortex ring, in the absence of both time-reversal and inversion symmetries; on loops linking this ring, the pseudospin winding can take on integer values (Fig. 1a shows the case of winding number 1). This gives rise to a toroidal smoke ring Fermi surface (Fig. 1a).

The model, besides describing a nodal ring spectrum (Fig. 1b) with an extended singularity, exhibits a new PCC corresponding to the pseudospin vortex ring - it is skyrmionic on both sides of the vortex ring, in the absence of a `fermion doubling problem' \cite{Nielsen81}. The new PCC implies a `maximal'  AHE for such a semimetal, as each planar cut through the Brillouin zone (planes parallel to the nodal ring plane) exhibits the same non-zero Chern number. The associated chiral Fermi arc, as a result, wraps around the \textit{full} surface Brillouin zone perpendicular to the nodal ring plane.

\begin{figure*}[!t]
\includegraphics[width=2.\columnwidth]{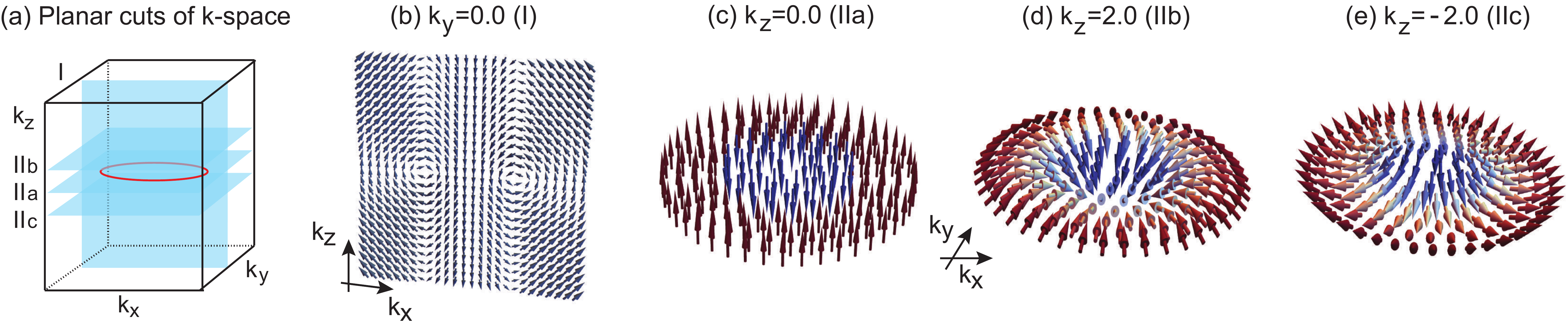}
\caption{(a) Different planar cuts of $\bf{k}$-space with fixed $k_y$ (labeled by I) and fixed $k_z$ (labeled by IIa, IIb, IIc). The red circle is the position of the nodal ring. (b)-(e) Pseudospin textures of the vortex ring Hamiltonian (with $k_0=2$) on various cuts of the $\bf{k}$-space. Away from rhw $k_z=0$ plane, (d) and (e), the pseudospin textures both carry a nonzero Pontryagin index $P(k_z)=1$.}
\end{figure*}
As a remarkable feature  of this  model, we find that this pseudospin vortex ring is connected to a monopole-anti-monopole pair of the Weyl semimetal--an example of a point defect pair annihilating without opening a gap.

Our model requires only a two-band construction. We also provide a lattice version based on a two-orbital tight-binding model. Its simplicity allows us to supply an analytical computation of its Landau level spectrum.

{\it The vortex ring (VR) model\,--- } Consider the two-band
\beq
H_{VR}(p_x,p_y,p_z)&=&-\f{1}{m_z}p_x p_z\sigma_x-\f{1}{m_z}p_y p_z\sigma_y\nn\\
&&\!\!\!\!\!\!\!\!\!\!\!\!+\biggl( \f{1}{2 m_r}(p_x^2+p_y^2-p_z^2)-\f{p_0^2}{2m_r} \biggr)\sigma_z,
\eeq
with Pauli matrices
$\bm{\sigma}$ acting on orbital/sublattice space without electron spin degeneracy. $p_0$ sets the radius of the nodal ring centered on the origin of the $p_z=0$ plane, while $p_0/m_r$ ($p_0/m_z)$ sets the Fermi velocities in (normal to) that plane, respectively.

We demonstrate the resulting pseudospin texture in the form of a toroidal magnetic field with quantized circulation around the axis of revolution (Fig. 1a) by the following non-perturbative procedure. First, in 2D, a massless Dirac Hamiltonian  in the $(p_x,p_z)$ plane with unit Fermi velocity can be written as $H_{Dirac}(p_x,p_z)=-p_z\sigma_x+p_x\sigma_z$. Here, the pseudospin winds an angle $+2\pi$ (giving the $\pi$ Berry phase) on a counter-clockwise circuit enclosing the Dirac point (Fig. 1a). Analogously, related to  a graphene bilayer, a 2D Hamiltonian with two vortices of equal winding (and a resulting $2\pi$ Berry phase) is given as $H_{bi}(p_x,p_z)=-(p_x p_z)/(m_z) \sigma_x+(p_x^2-p_z^2-p_0^2)/(2 m_r) \sigma_z$ (see e.g., Refs. \cite{Gail11,Mon09}). The global $2\pi$ Berry phase is distributed among two unit vortices at $(p_x,p_z)^{\pm}=(\pm p_0,0)$.

On rotating $H_{bi}(p_x,p_z)$ around the $p_z$ axis, the two isolated Dirac nodes trace out a circular nodal line in $\bm{k}$-space, resulting in the vortex ring Hamiltonian (1) with unit winding around the axis of revolution. With this procedure, a sequence of vortex ring Hamiltonians with higher winding can also be generated, see the Supplemental Material \cite{supp}.

The resulting energy spectrum exhibiting a nodal ring of radius $k_0$  is given by (setting $m_r=m_z=\hbar=1$, $\bm{p}=\bm{k}$)
\beq
E_{\pm}=\pm\sqrt{(k_r^2-k_z^2-k_0^2)^2/4+k_r^2 k_z^2}
\eeq
with the radial wave vector $k_r\equiv(k_x^2+k_y^2)^{1/2}$ (Fig. 1b).

The stability of the nodal ring arises from  a particular `mirror reflection' symmetry: a reflection with respect to the $z=0$ mirror plane, combined with an opposite parity of the two orbitals under such a transformation. The Bloch Hamiltonian $H(\bm{k})=\bm{h}(\bm{k})\cdot \bm{\sigma}$ thus transforms as $\bm{h}(\bm{k})\rightarrow (-h_x(k_x,k_y,-k_z),-h_y(k_x,k_y,-k_z),h_z(k_x,k_y,-k_z))$. The gaplessness of the nodal ring spectrum  is then protected by the mirror symmetry: $h_x=h_y=0$ on the mirror plane. The model explicitly breaks time-reversal and inversion symmetry \cite{foot1}.

As an aside, we mention another type of nodal-ring Hamiltonian $H=(k_r^2-k_0^2)\sigma_x+k_z\sigma_y$ (see e.g., Refs. \cite{Weng15b,Fang15,Mullen15,Kim15}) requiring time-reversal and inversion symmetries. The absence of the $\sigma_z$ component results in a planar pseudospin configuration throughout $\bm{k}$-space. Even though this does carry a Berry phase feature, the  topological features discussed below for $H_{VR}$ are absent.

{\it The pseudospin Skyrmion\,--- }
We characterize the global characteristic of the pseudospin vortex ring on different planar cuts of $\bm{k}$-space, labeled as I and IIa-IIc in Fig. 2a. First, by construction the pseudospins on the $k_y=0$ plane (I) are strictly planar with two vortices of equal winding (Fig. 2b). Second, on different $k_z$ planes (IIa-IIc) the pseudospins develop a full skyrmion structure when $k_z\neq 0$ \cite{Volovik03} (Figs. 2c-e), see Supplemental Material \cite{supp}. Note that the  sign of the Pontryagin index (Skyrmion number) of the mapping from the $k_x$-$k_y$ plane (with the `boundary points' at large $k_r$ identified) to the Bloch sphere (defined for the pseudospins) is independent of $k_z$, despite the different way the pseudospins wrap around the origin for $k_z$ of different sign (Figs. 2d,e).

 \begin{figure}
\begin{center}
\includegraphics[width=8.5cm]{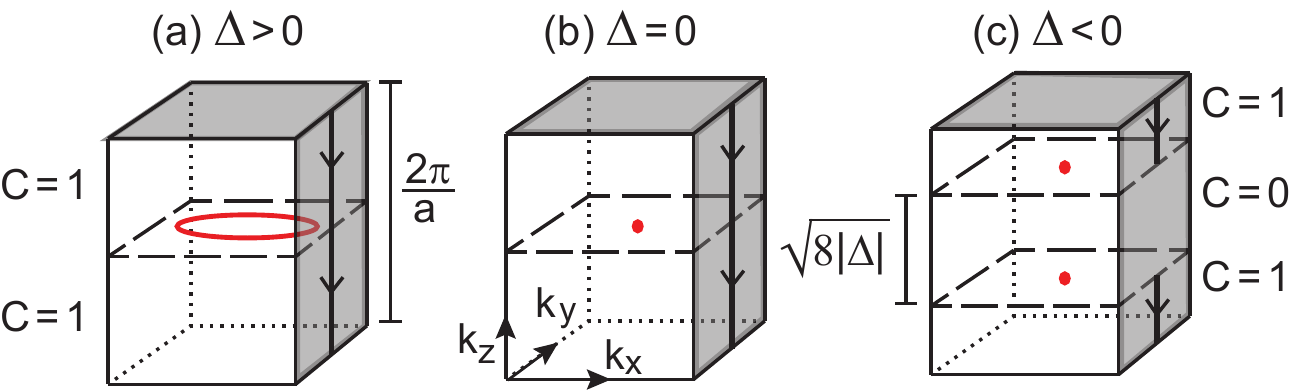}
\end{center}
\caption{The planar Chern composition (PCC) rule for the two pseudospin defects and the associated chiral Fermi arcs: (a) The nodal ring (red circle) lies on the $k_z=0$ plane in the 3D BZ. A pseudospin vortex ring exhibits $C=1$ on both sides of the singularity plane. (b) At the critical value $\Delta=0$ the nodal ring shrinks to a point (red dot). (c) The appearance of two Weyl nodes (two red dots) with a change in PCC, resulting in the opening of a region of width $\sqrt{8|\Delta|}$ with $C=0$. On the surface Brillouin zone (shaded region), the locus of the zero-energy chiral Fermi arc are shown (directed bold line).}
\end{figure}
The family of 2D Hamiltonians, $H_{k_z}^{2D}$, parametrised by  $k_z$ via $H_{VR}=H_{k_z}^{2D}(k_x,k_y)$, represent 2D Chern insulators with Chern number $C(k_z)=1$ for $k_z\neq 0$. Thus we have a new PCC where $C=1$ on \textit{both} sides of the vortex ring (Fig. 3a). Contrast this with  the  change  $C=0\rightarrow 1$ as the singularity of a pseudospin monopole is crossed. This turns out to be crucial in the following.

{\it Tight-binding realization and AHE\,--- }
To discuss the 3D intrinsic AHE, we provide a tight-binding model for the vortex ring Hamiltonian. We introduce two lattice constants, $a$ and $b$, for the $z$- and $x,y$- directions, respectively. The nodal ring diameter $2 k_0$ is then measured as a fraction of the planar reciprocal lattice $\alpha 2 \pi/b$, for $0<\alpha<1$. It is given as $H_{TB}=\bm{h}(\bm{k})\cdot \bm{\sigma}$ with
\beq
h_x(\bm{k})&\propto&-(a b)^{-1}\, \sin(k_x b)\sin(k_z a/2),\nn\\
h_y(\bm{k})&\propto&-(a b)^{-1}\, \sin(k_y b)\sin(k_z a/2),\nn\\
h_z(\bm{k})&\propto&b^{-2}\,( 1-\cos(k_x b)-\cos(k_y b)+\cos(\alpha \pi))\nn\\
&&-\tilde{\alpha} a^{-2}(1-\cos(k_z a)),
\eeq
where $0<\tilde{\alpha}<1$ with the full expressions given in the Supplemental Material \cite{supp}. Expanding $H_{TB}$ close to the nodal ring $k_z=0$, $k_r=\alpha \pi/b \,(=k_0)$ yields Eq. (1).

\begin{figure}
\begin{center}
\includegraphics[width=8.2cm]{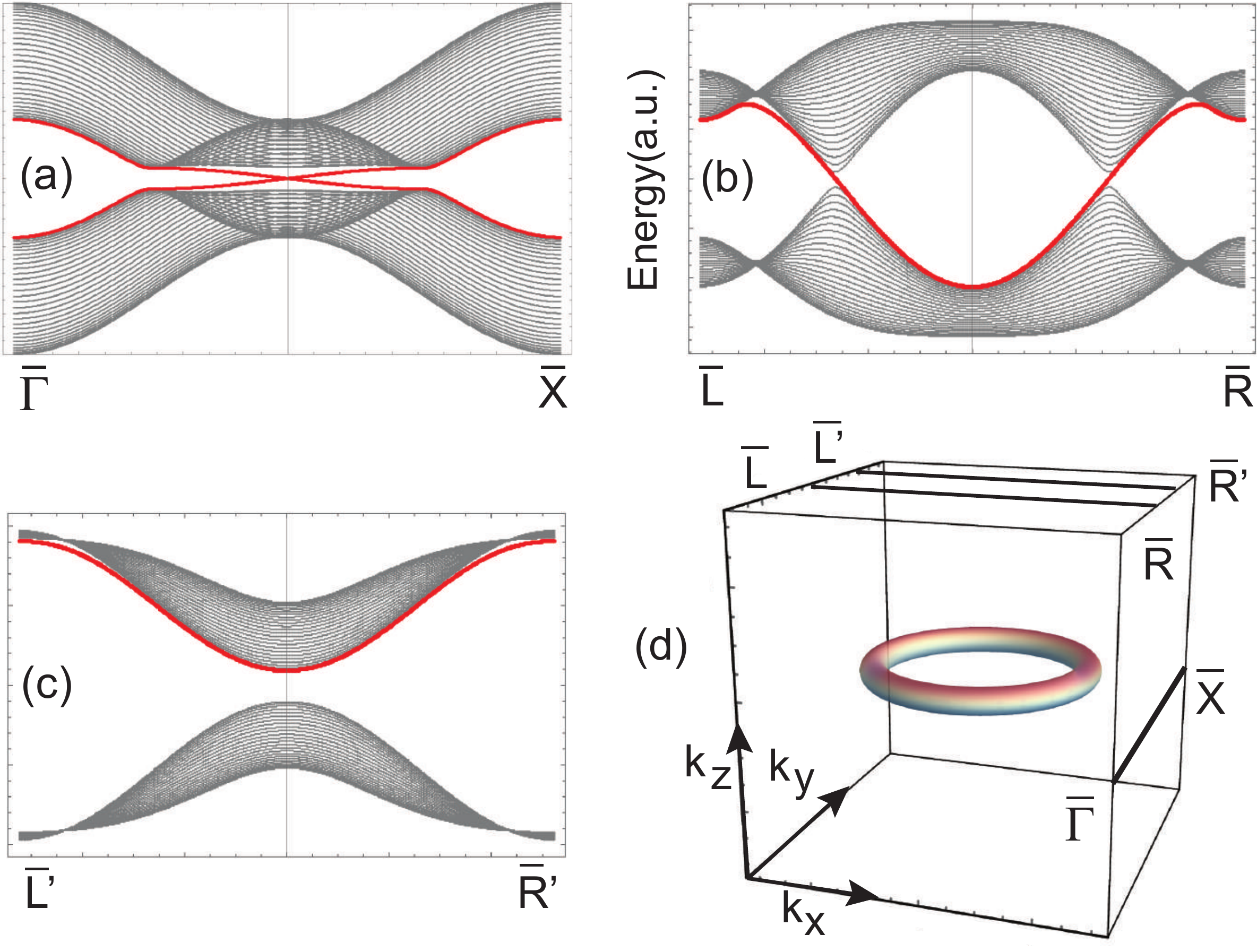}
\end{center}
\caption{(a)-(c) Two-dimensional surface band structures on planes perpendicular (a) and parallel (b,c) to the nodal ring plane (defined in (d)). Surface states are shown in red \cite{foot3}. (d) Normalized Fermi surface Berry curvature $\bm{n}(\bm{k}_F)\cdot \bm{\Omega}(\bm{k}_F)\in[-1 \tr{ (red)},1 \tr{ (blue)}\,]$ in the topological nodal ring phase with a fixed Fermi energy $E_F=0.5$. $\bm{n}$ is the unit normal vector on the FS, $\bm{\Omega}$ is the Berry curvature and $\bm{k}_F$ the Fermi wave vector ($m_r=m_z=\hbar=1$). }
\end{figure}
This describes a two-orbital model  on a tetragonal lattice. The form factor $\sin(k_z a/2)$ in the inter-orbital hopping term $h_{x,y}$ indicates a two-site basis along the $z$-direction, see Supplemental Material \cite{supp}. The energy spectrum exhibits one nodal ring per Brillouin zone (Fig. 3a).

Contrary to the fermion doubling in lattice realizations of Weyl/Dirac fermions \cite{Nielsen81}, there is no analogous issue here, i.e., there is no topological obstruction to having only  one nodal ring in the full Brillouin zone.

The non-zero Chern number of $H_{k_z\neq 0}^{2D}$ implies the existence of 1D chiral edge states on the boundary of finite systems. On the surface Brillouin zone parallel to $k_z$, the zero-energy chiral surface states form a Fermi arc \cite{Wan11, Yang11, Hosur12,Haldane14,foot2}. For the vortex ring phase, in fact, it wraps around the \textit{full} surface BZ, see Figs. 3a-b. This is confirmed by numerically solving a finite $H_{TB}$, see Fig. 4a, where localized chiral states cross the bulk energy gap as required by non-trivial band topology. As a physical consequence, this amounts to $H_{VR}$ describing a novel kind of 3D topological semimetal: it has a gapless nodal ring in the bulk and an intrinsic, `maximal' anomalous Hall effect with a Hall conductivity $\sigma_{xy}^{3D}=(e^2/2\pi h)(2 \pi/a)$, where $(2 \pi/a)$ is the magnitude of the primitive reciprocal vector perpendicular to the nodal ring plane.

{\it Transition to a Weyl semimetal\,--- }
The model Hamiltonian (1) also describes the Weyl phase. As $\Delta\equiv p_0^2/2m_r$ is swept towards 0, the nodal ring shrinks,
turning  into a point at $\Delta=0$, whereafter two Weyl nodes appear at $\bm{k}_W^{\pm}=(0,0,\pm\sqrt{2 m_r |\Delta|/\hbar^2})$ for $\Delta<0$ (Fig. 3). The low-energy Hamiltonian around $\bm{k}_W^{\pm}$ is given by $H_{VR}\approx \pm \sqrt{2 m_r |\Delta|} ((p_x/m_z)\sigma_x+(p_y/m_z)\sigma_y+(p_z/m_r)\sigma_z)$, describing a monopole-anti-monopole pair.

This is remarkable because upon annihilation of such a pair (consider sweeping $\Delta$ in an opposite direction), one might have
expected an energy gap to open \cite{Klimkhamer05}. Here we explicitly show that the $H_{VR}$ offers a much richer scenario:
a continuous change from a vortex ring to a monopole-anti-monopole pair, remaining gapless throughout.

For the intrinsic AHE, as $\Delta \rightarrow 0^+$ the Hall conductivity retains its  value of the nodal ring phase; for $\Delta<0$, $\sigma_{xy}^{3D}=(e^2/2\pi h) (2\pi/a- \sqrt{8 m_r |\Delta|/\hbar^2})$. The latter is the known intrinsic unquantized AHE of the Weyl semimetal \cite{Klimkhamer05,Yang11,Burkov11a}. In this sense, the model Hamiltonian describes, on the one hand, a topological nodal ring phase $(\Delta>0)$ with its maximal AHE and on the other hand, it is `adiabatically' connected to the Weyl phase $(\Delta<0)$ with two Weyl nodes.

As an aside, we mention that this AHE in the topological nodal ring phase is robust for a smoke ring Hamiltonian of the more general type. Specifically, when the mirror symmetry is removed,  the nodal  ring is gapped; however, the PCC rule for vortex ring remains intact (see Supplemental Material \cite{supp}).

Another quantity of interest is the Fermi surface (FS) Berry curvature \cite{Haldane04,Haldane14,GM15}. The FS topology, at small, finite electron density, evolves from a torus-shaped geometry (genus 1), via a sphere (genus 0) to two disconnected closed sheets, thus experiencing multiple Lifshitz transitions. In the nodal ring phase, the FS Berry curvature is non-zero everywhere except on the $k_z=0$ plane,  see Fig. 4d - thus, possessing the basic ingredient for `nonlocal transport' \cite{Parameswaran14,Gorbachev14}. 

\begin{figure}
\begin{center}
\includegraphics[width=7.5cm]{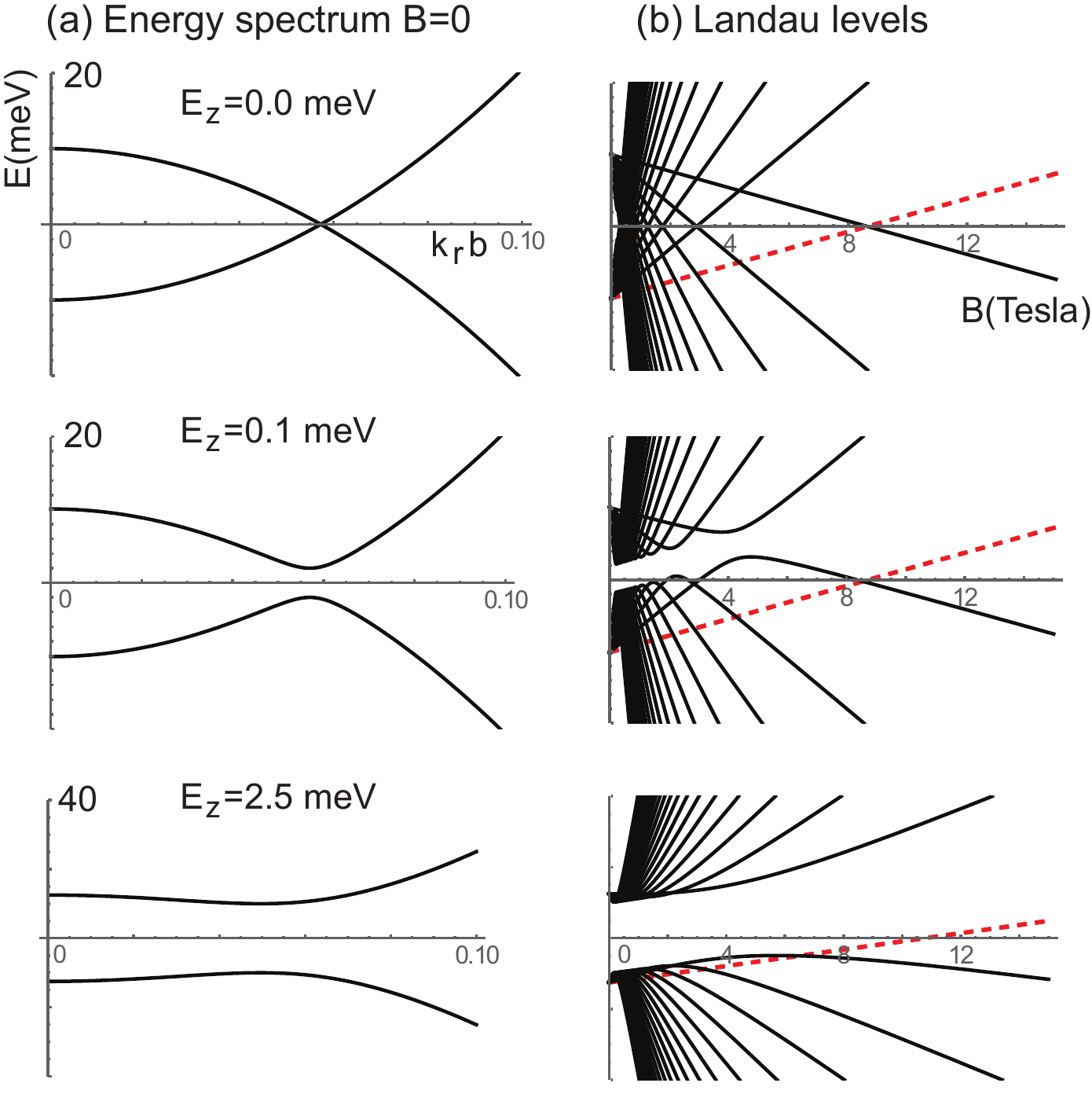}
\end{center}
\caption{Magnetic field dependence of the Landau levels for various values of $E_z=p_z^2/2m$. On the left panel is the energy spectrum in the absence of the magnetic field. On the right panel, the black (dashed, red) curves correspond to $E_n$ ($E_0$). We use crystal parameters $b=5$\AA, $m=0.05 m_e, \Delta=10\tr{meV}$.}
\end{figure}

{\it Landau level structure\,--- }
As a basis for the study of magnetotransport properties, we now turn to the quantum mechanical Landau level (LL) problem. The two-band vortex ring model permits a fully analytic solution, and reveals an anomalous LL state.

From the Hamiltonian (1), we use the minimal substitution for a magnetic field $\mbf{B}=B \hat{z}$, $B>0$, and promote the relevant conjugate variables to operators $(p_x,p_y)\rightarrow(\hat{p}_x - e\hat{A}_x, \hat{p}_y-e\hat{A}_y)$, with $(\hat{A}_x,\hat{A}_y)=(-B \hat{y}/2,B \hat{x}/2)$. Since the momentum in the $z$-direction remains a good quantum number, the problem decomposes into a family of 2D ones parametrized by $p_z$. Introducing the ladder operators $\hat{a}=\sqrt{2}(\partial_{\bar{z}}+z/4)$, $\hat{a}^\dag=\sqrt{2}(-\partial_{z}+\bar{z}/4)$, with $z=x+i y\, (\bar{z}=x-i y)$, such that $[\hat{a},\hat{a}^\dag]=1$, we arrive at
\beq
H=\bem \hat{a}^\dag \hat{a} +\f{1}{2}-\delta_1 & -i\delta_2 \hat{a}^\dag\\
i\delta_2 \hat{a} & -(\hat{a}^\dag \hat{a} +\f{1}{2}-\delta_1)
\eem
\eeq
where $\delta_1=(E_z+\Delta)/\epsilon_B,\delta_2=\sqrt{4E_z/\epsilon_B} \,\tr{sgn}(p_z)$ and $E_z=p_z^2/2m$ with $\epsilon_B=\hb eB/m$.
The Landau levels are (see Supplemental Material \cite{supp} for details)
\beq
&&E_{n=0}=\f{1}{2}\f{\hbar e}{m}B-(E_z+\Delta),\nn\\
&&E_{n}^{\pm}=\f{1}{2}\f{\hbar e}{m}B\pm[n^2\f{\hbar^2 e^2}{m^2}B^2+2n\f{\hbar e}{m}B(E_z-\Delta)\nn\\
&&\tr{\ \ \ \ \ \ \ }+(E_z+\Delta)^2]^{1/2}\tr{\ \ for }n\geq1,
\eeq
plotted in Fig. 5 for various $p_z$. As a reference for the quantization pattern, the energy spectra in the absence of the magnetic field are plotted on the left.

A most salient feature of the LL is the $n = 0$ eigenstate with the `wrong' slope (dashed (red) line in Fig. 5). It plays a role similar to the zero-energy LL of graphene, occupying only one of the two sublattices $(|0\rangle, 0)^T$. Unlike the latter, the $n=0$ state has a positive slope despite being a holelike state. As a result, it transmutes into a particlelike state at sufficiently large fields. This is reminiscent of the anomalous LL spectrum of a 2D spin Hall insulator found in a quantum well semiconductor \cite{Konig07}, reflecting the underlying 2D Chern insulating character.

Finally, to complement the quantum results, we apply Onsager's relation to obtain the semiclassical LL given by
$E^{semicl}_n=\pm [\bigl(\hbar eB(n+\gamma_{\lessgtr})/m-\Delta+E_z\bigr)^2+4 \Delta E_z]^{1/2}$, with an undermined index $\gamma_{\lessgtr}$ corresponding to the two cyclotron orbits on the Fermi surface. In the large-$n$ (semiclassical) limit, both the semiclassic analysis and the quantum result agree up to the $\mathcal{O}(n^0)$ \cite{Fuchs10}. The matching condition yields $\gamma_{\lessgtr}=\pm\f{1}{2}$ indicating a trivial Berry phase, as expected for cyclotron orbits in a coupled-parabolic-band problem.

{\it Summary and outlook\,--- } We have presented a 3D semimetal that exhibits a novel intrinsic anomalous Hall effect. This follows
from the observation of a new topological character in the band structure arising from a nodal line system with a vortex ring singularity.
Requiring only a tetragonal tight-binding model satisfying a mirror symmetry, and strictly local hopping, it would appear not to be entirely unreasonable to hope for an actual material \cite{Weng16s} or cold atom realisation \cite{Dub15,Xu16,DZhang16}. Interesting open questions include investigating the phase transition across the topological nodal ring and Weyl phases, in the spirit of the Weyl semimetal-insulator transition in Refs.~\cite{Dora13,Yang14}, and of course the effect of interactions more generally \cite{Roy16,Sur16}. Moreover, complete classification and study of the connection between different momentum space singularity structures remains a
largely unexplored subject.
\medskip

\acknowledgments
We thank Jean-No\"{e}l Fuchs, Titus Neupert, Zhong Wang and two anonymous referees for useful discussions and comments on the Letter. This work was in part supported by Tsinghua University Initiative Research Programme, the Thousand Young Talents Program of China (L.-K. L.) and DFG under Grant No. SFB 1143 (R.M.).

\widetext

\newpage
\section{Supplementary material}
\section{Vortex ring Hamiltonian of arbitrary winding}
\setcounter{equation}{0}
We outline the procedure to obtain the pseudospin vortex ring Hamiltonian of arbitrary winding. Starting with a 2D graphene-bilayer-like Hamiltonian, $H_{bi}(p_x,p_z)$ (see main text), with purely planar pseudospins, the out-of-plane pseudospins are generated by rotating the Hamiltonian around the $p_z$-axis. The procedure is analogous to rotating a $2$-vector on the plane: $p_x$ becomes the radial component $p_x\rightarrow \sqrt{p_x^2+p_y^2}$ with the azimuthal angle $\phi= \arctan(p_y/p_x)$, $0\leq \phi < 2\pi $. After this substitution, we  multiply the $\sigma_x$ component of the bilayer Hamiltonian with the projection factors $\cos \phi$ and $\sin \phi$ to obtain the respective
new pseudospin components in the $\sigma_{x,y}$ axes: $H_{VR}(\bm{p})=\bm{h}(\bm{p})\cdot\bm{\sigma}$ with
\beq
h_x(\bm{p})&=&-\f{1}{m_z}\sqrt{p_x^2+p_y^2}\,\,p_z\cos \phi,\nn\\
h_y(\bm{p})&=&-\f{1}{m_z}\sqrt{p_x^2+p_y^2}\,\,p_z\sin \phi,\nn\\
h_z(\bm{p})&=&\f{1}{2m_r}(p_x^2+p_y^2-p_z^2)-\f{p_0^2}{2 m_r}.
\eeq
With $\sqrt{p_x^2+p_y^2}\cos\phi = p_x$, $\sqrt{p_x^2+p_y^2}\sin\phi = p_y$,  one arrives at the Hamiltonian (1).

\begin{figure}[!b]
\begin{center}
\includegraphics[width=5cm]{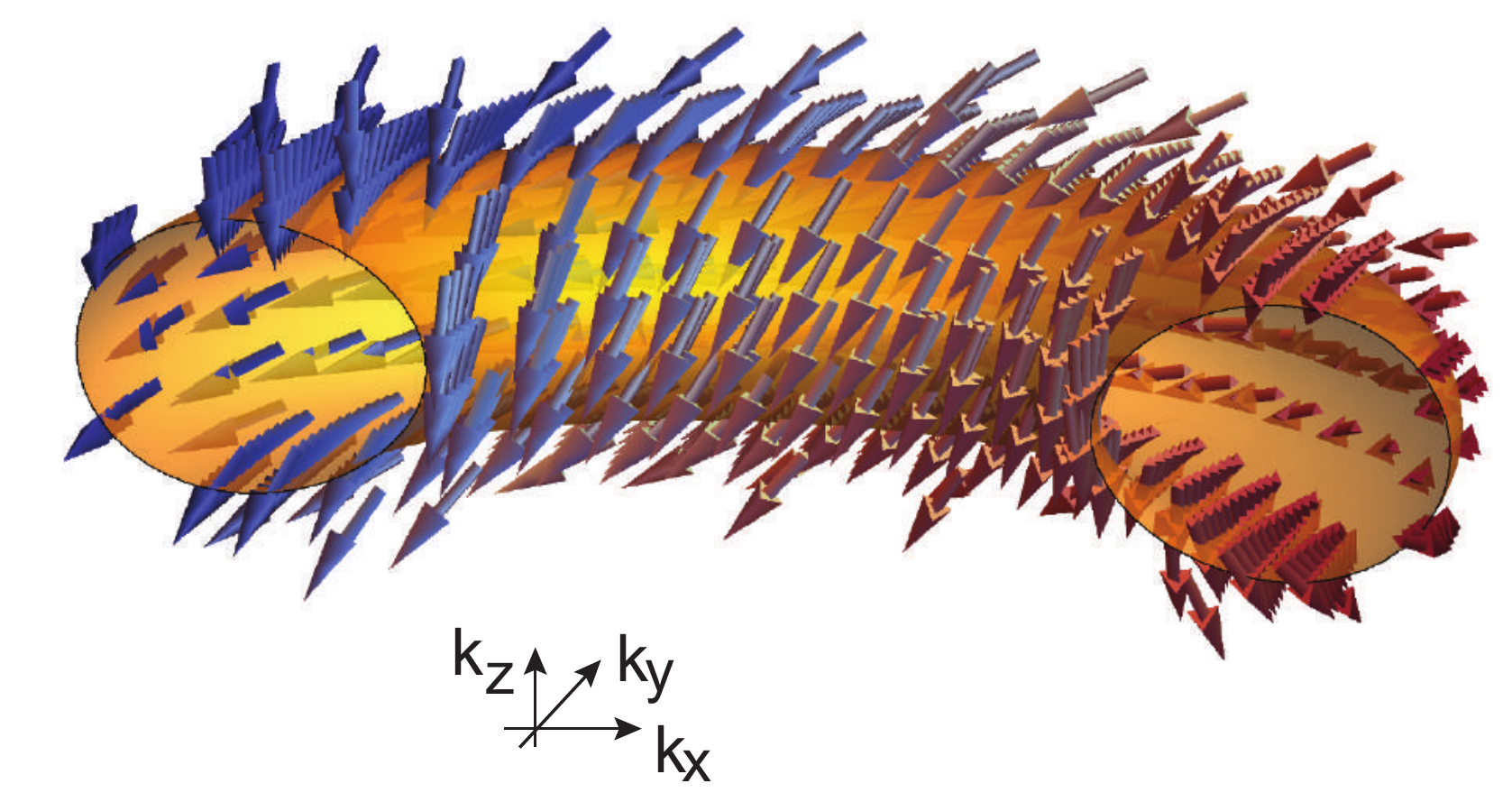}
\end{center}
\caption{General smoke ring pseudospin structure shown on the toroidal Fermi surface close to the nodal ring.}
\end{figure}

The same procedure can be generalized to obtain vortex ring Hamiltonian with arbitrary winding. Here we give the results of vortex ring Hamiltonian with winding 2 and 3 (setting $m_r=m_z=\hbar=1$, $\bm{p}=\bm{k}$). Starting with the 2D graphene-bilayer-like Hamiltonians with two vortices sharing equally the global $4\pi$ and $6\pi$ Berry phases:
\beq
H_{bi}^{4\pi}(k_x,k_z)&=&[-4(k_x^2-k_0^2)k_x k_z+4 k_x k_z^3]\sigma_x+[(k_x^2-k_0^2)^2+k_z^4-6 k_x^2 k_z^2]\sigma_z;\nn\\
H_{bi}^{6\pi}(k_x,k_z)&=&[-6(k_x^2-k_0^2)^2 k_x k_z+20 k_x^3 k_z^3-6 k_x k_z^5]\sigma_x+[(k_x^2-k_0^2)^3 -15 (k_x^2-k_0^2)k_x^2 k_z^2+15 k_x^2 k_z^4- k_z^6]\sigma_z,
\eeq
by rotation we obtain the vortex ring Hamiltonians:
\beq
H_{VR,2}(\bm{k})&=&[-4(k_r^2-k_0^2)k_x k_z+4k_x k_z^3]\sigma_x+[-4(k_r^2-k_0^2)k_y k_z+4k_y k_z^3]\sigma_y\nn\\
&&+[(k_r^2-k_0^2)^2+ k_z^4-  6  k_r^2 k_z^2]\sigma_z;\\
H_{VR,3}(\bm{k})&=&[-6(k_r^2-k_0^2)^2k_x k_z+20 k_x k_r^2 k_z^3-6 k_x k_z^5]\sigma_x\nn\\
&&+[-6(k_r^2-k_0^2)^2k_y k_z+20 k_y k_r^2 k_z^3-6 k_y k_z^5]\sigma_y\nn\\
&&+[(k_r^2-k_0^2)^3-15 (k_r^2-k_0^2)k_r^2k_z^2+15 k_r^2 k_z^4-k_z^6]\sigma_z.
\eeq

\subsection{General smoke-ring Hamiltonian}
We discuss a more general smoke ring Hamiltonian $H_{SR}$ which does not exhibit the mirror symmetry protecting the nodal line which is thus
gapped out. We begin with a 2D \textit{gapped} Hamiltonian $H(k_x,k_z)= -k_x k_z \sigma_x+k_x \sigma_y+((k_x^2-k_z^2)/2-\Delta)\sigma_z$ for $\Delta>0$. Rotating the Hamiltonian around the $k_z$-axis as before, we obtain
\beq
H_{SR}(\bm{k})=(-k_x k_z-k_y) \sigma_x+(-k_y k_z+k_x)\sigma_y+((k_x^2+k_y^2-k_z^2)/2-\Delta)\sigma_z.
\eeq
From Fig. A1, we see that the gapped energy spectrum leads to an additional twist of the pseudospin texture on the toroidal Fermi surface, c.f. Fig 1a. The PCC rule, however, remains the same as for the vortex ring Hamiltonian studied in the main text. As $\Delta>0$ is swept towards 0, the gap closes at a single point. For $\Delta<0$, two Weyl nodes appear. This can serve as an explicit low-energy model that describes the annihilation of two Weyl nodes and resulting in a 3D quantum anomalous Hall insulator phase \cite{Burkov11ap}.

\section{Pontryagin index, Berry curvature, Chern number}
By parameterizing the vortex ring Hamiltonian as $H_{VR}=\bm{h}(\bm{k})\cdot \bm{\sigma}=|\bm{h}(\bm{k})|\, \bm{n}(\bm{k})\cdot \bm{\sigma} $, the Pontryagin index for a given $k_z$ is given as $P(k_z)=(1/4\pi)\int dk_x dk_y \,\bm{n}\cdot (\partial_{k_x}\bm{n}  \times \partial_{k_y}\bm{n})=1$ for $k_z\neq 0$. The normalized pseudospins $\bm{n}(\bm{k})$ at large $k_r$ are identified as the same point $\bm{n}=(0,0,1)$.

The Chern number for the vortex ring Hamiltonian can be evaluated to give $C(k_z)=(1/2\pi)\int_{1BZ} dk_x dk_y \Omega_{k_x k_y} =1$ for $k_z\neq 0$, where $\bm{\Omega}(\bm{k})=\nabla_{\bm{k}} \times \bm{A} (\bm{k})$ and $ \bm{A} (\bm{k})=i \langle n (\bm{k})|\partial_{\bm{k}}| n (\bm{k}) \rangle$ are the Berry curvature and the Berry connection, respectively, using the lower-band eigenstates of  $H_{VR} | n (\bm{k}) \rangle = -|\bm{h}(\bm{k})|| n (\bm{k}) \rangle$ with periodic boundary condition.

\begin{figure}
\begin{center}
\includegraphics[width=9.3cm]{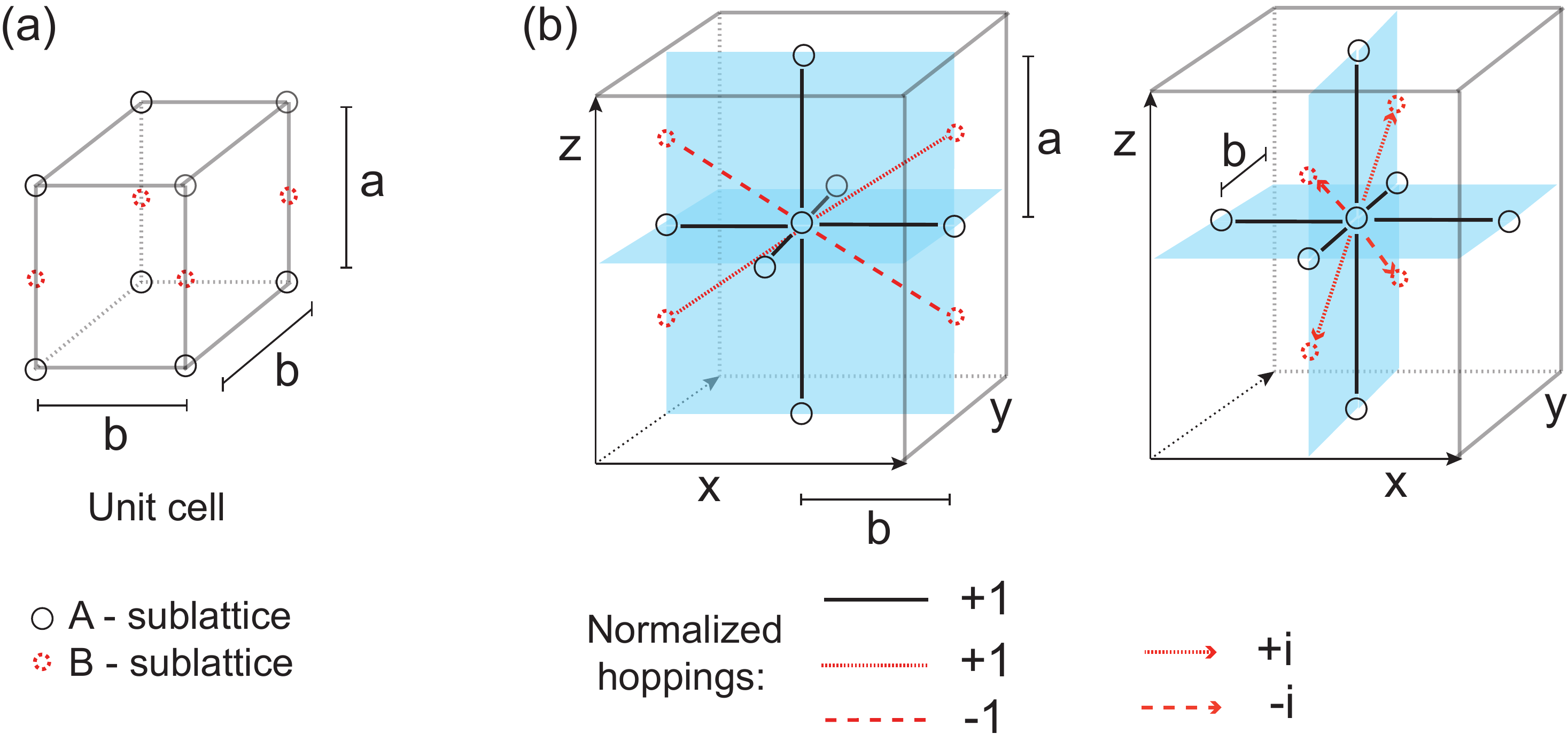}
\end{center}
\caption{(a) Unit cell structure for the tight-binding realization. (b) Real space visualization of the hoppings described by $H_{TB}$ on the $x$- (left), and the $y$-plane (right), respectively.}
\end{figure}
\section{Tight-binding realizations}
The full expression corresponding to Eq. (4) is given as $H_{TB}=\bm{h}(\bm{k})\cdot \bm{\sigma}$ with
\beq\label{TB}
h_x(\bm{k})&=&-\f{2\hbar^2}{m_z a b}\f{\alpha \pi}{\sin(\alpha \pi)}\, \sin(k_x b)\sin(k_z a/2)\nn\\
h_y(\bm{k})&=&-\f{2\hbar^2}{m_z a b}\f{\alpha \pi}{\sin(\alpha \pi)}\, \sin(k_y b)\sin(k_z a/2)\nn\\
h_z(\bm{k})&=&\f{\hbar^2}{m_r b^2}\f{\alpha \pi}{\sin(\alpha \pi)}\,\biggl( 2-\cos(k_x b)-\cos(k_y b)\nn\\
&&+\cos(\alpha \pi)-1 \biggr)-\f{\hbar^2}{m_r a^2}(1-\cos(k_z a))
\eeq
and illustrated in Fig. A2. By expanding $H_{TB}$ around the nodal ring $k_z=0$, $k_r=k_0=\alpha \pi/b$, the low energy nodal ring Hamiltonian (1) is recovered. Note that the nodal ring is not perfectly circular anymore in the tight-binding model: it inherits the underlying square symmetry of the lattice realization.

In Fig. A3, we show the pseudospins in the $k_y=0$ plane covering three Brillouin zones. In the first Brillouin zone (marked with a box) we recover the full feature of the low energy Hamiltonian (1), see Fig. 2b. In the second B.Z., the pseudospins consist of an anti-vortex-anti-vortex pair instead. Due to the sublattice structure along $z$-direction, the Bloch Hamiltonian generally has a different periodicity (or no periodicity) from the Brillouin zone periodicity \cite{Lim15}.

Similar to the apparent differences between the skyrmion structures discussed for Fig. 2d and 2e but resulting in the same winding property, the planar Chern number (on the $k_x,k_y$ plane) is a constant $C=1$ throughout, except at the gap closing planes $k_z=0, \pm2\pi/a, \pm4\pi/a, \ldots,$ where $C$ is not defined. This is the new PCC rule across a vortex ring singularity.

To study the surface states evolution as the system turns from the vortex ring phase into the Weyl semimetal phase, we numerically solve two instances of the tight-binding model $H_{TB}$ with finite extent in the x-direction, giving surface bandstructures parallel to $k_z$, see Fig. A4. In the vortex ring phase, chiral surface states exist for all $k_z\neq 0$, giving rise to the fully extended Fermi arc. In the Weyl semimetal phase, chiral surface states exist for $k_z$-plane with unit Chern number, resulting in an open-ended Fermi arc. A summary of the Fermi arc structure is shown in Fig. 3 in the main text.

\begin{figure}
\begin{center}
\includegraphics[width=7.5cm]{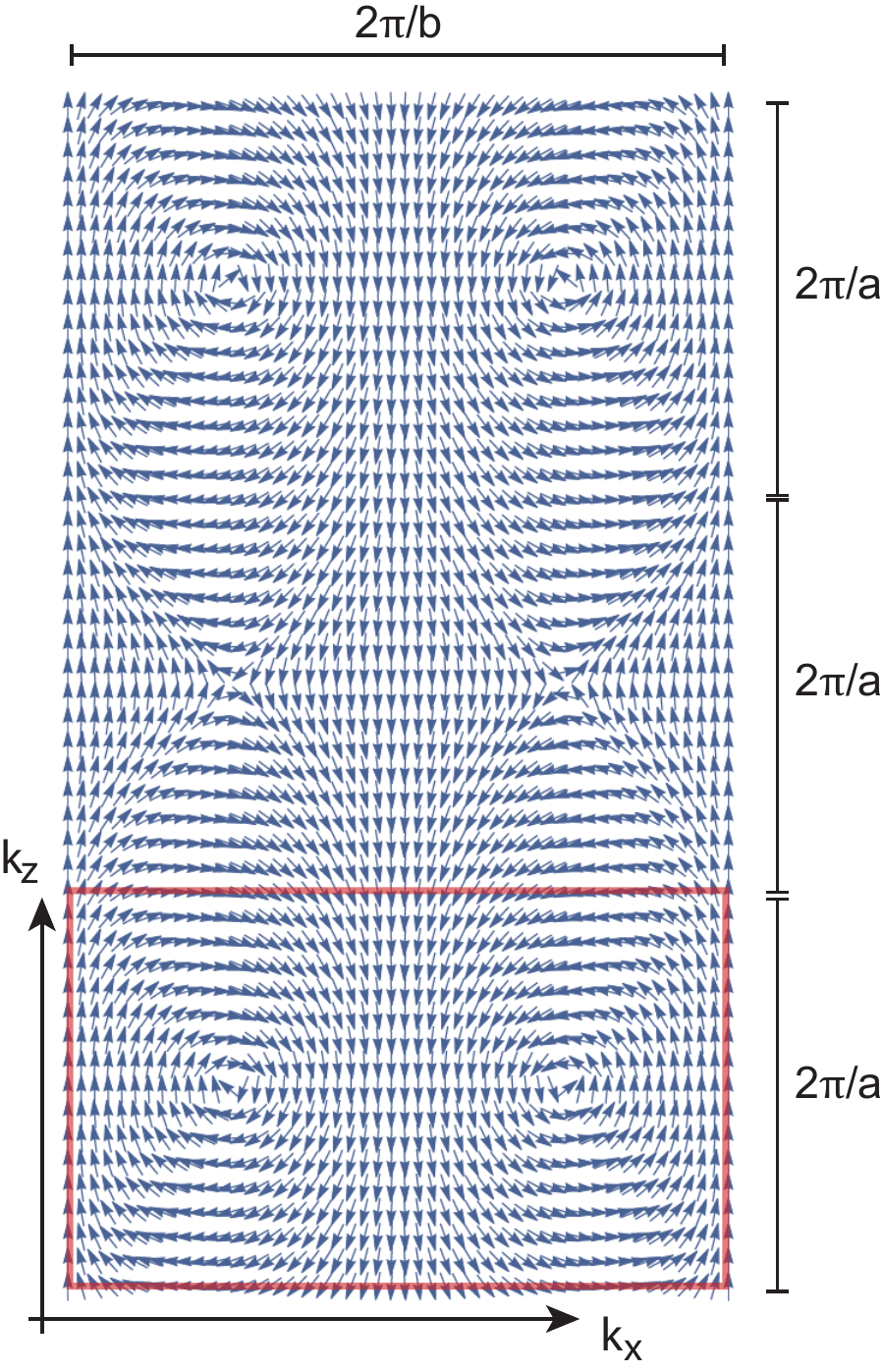}
\end{center}
\caption{Pseudospin structure of $H_{TB}$ (with $k_y=0$) covering three Brillouin zones in the $k_z$ direction (setting $\hbar=m_z=m_r=1$, $b=0.6$, $a=1$). The area mark with a box indicates the first Brillouin zone.}
\end{figure}

\begin{figure}
\begin{center}
\includegraphics[width=15cm]{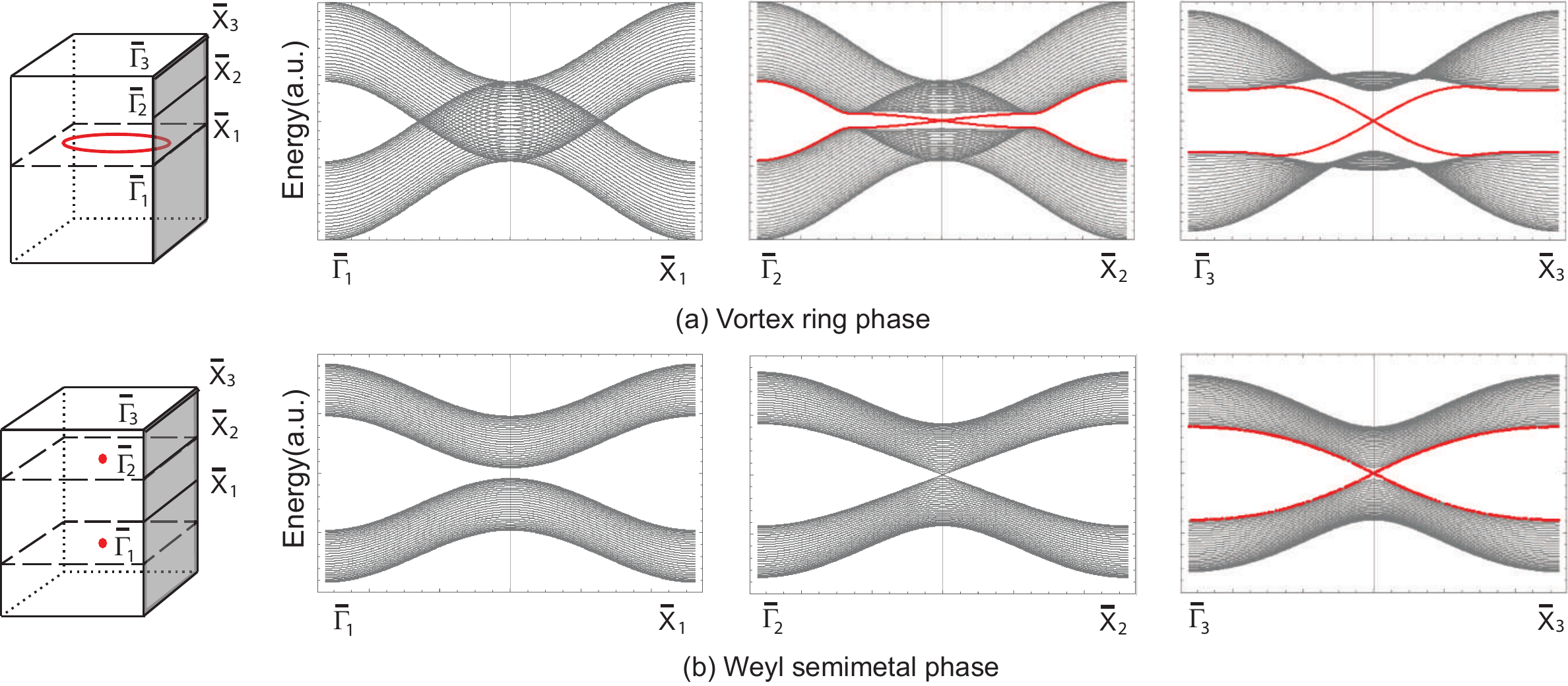}
\end{center}
\caption{Two-dimensional surface bandstructures along the various quasimomentum range $\bar{\Gamma_i}-\bar{X_i}$, for $i=1-3$: (a) in the vortex ring phase (b) the Weyl semimetal phase. Energy levels of surface states are indicated in red.}
\end{figure}

\section{Solution to the Landau levels}
In this section, the details of the calculation for the Landau level problem of the vortex ring Hamiltonian are outlined. We first show the limit when the two bands are uncoupled, i.e., when $E_z=0$, and then show the general case. The former reproduces the expected result of two inverted quadratic bands, to serve as a reference point where simple results are known.

\subsection{$E_z=0$ case}
In the limit $E_z=0$, the two quadratic bands are uncoupled. The Hamiltonian in a magnetic field simplifies to
\beq
H/\epsilon_B=\bem \hat{a}^\dag \hat{a} +\f{1}{2}-\delta_1 & 0\\
0 & -(\hat{a}^\dag \hat{a} +\f{1}{2}-\delta_1)
\eem
\eeq
and the eigenvalues are
\beq
E_n^{\pm}=\pm (n+\f{1}{2}-\delta_1)\epsilon_B
\eeq
with  eigenfunctions $(|n \rangle,0)^T$ and $(0,|n \rangle)^T$ for $n\geq 0$, where $\hat{a}^\dag \hat{a} |n\rangle=n|n\rangle$. The results are those of the Landau levels of two quadratic bands, inverted with respected to each other, and shifted by the energy $\Delta=\delta_1 \epsilon_B$.

\subsection{General case}
We start with the LL Hamiltonian given by (see Eq. (5) in the main text)
\beq
H/\epsilon_B=\bem \hat{a}^\dag \hat{a} +\f{1}{2}-\delta_1 & -i\delta_2 \hat{a}^\dag\\
i\delta_2 \hat{a} & -(\hat{a}^\dag \hat{a} +\f{1}{2}-\delta_1)
\eem.
\eeq
In contrast to the corresponding Landau level problem in graphene, ``squaring" the Hamiltonian does not render it diagonal. Instead, we seek the solution using an ansatz formed by the basis for the two-level problem $\{ (|n\rangle,0)^T,(0,|n\rangle)^T\}$, $\{ (|n\rangle,|m\rangle)^T,(|n\rangle,|m\rangle)^T\}$, for $n,m\geq 0$, with $\hat{a}^\dag \hat{a} |n\rangle=n|n\rangle$.
The $n=0$ Landau level solution is given by
\beq
H \bem |0\rangle\\ 0 \eem = E_{n=0}  \bem |0\rangle\\ 0 \eem
\eeq
with $E_{n=0}=(1-\delta_1)\epsilon_B$. For $n\geq 1$ the solution is obtained by solving
\beq
H \bem \cos(\theta_n/2) |n\rangle  \\  e^{i\phi_n} \sin(\theta_n/2) |n-1\rangle \eem = E_n  \bem \cos(\theta_n/2) |n\rangle  \\  e^{i\phi_n} \sin(\theta_n/2) |n-1\rangle \eem
\eeq
with $\phi_n, \theta_n$ parameterizing the $n$-th spinor eigenfunction of the ``two-level" problem. By demanding self-consistency, we obtain $\phi_n=\pm \pi/2 \,\tr{sgn} (p_z) $ and
\beq
\tan\f{\theta_n}{2}= \f{\sqrt{(n-\delta_1)^2+n \delta_2^2 }\pm (\delta_1-n)}{|\delta_2|\sqrt{n}},\tr{\ \ }n\geq1,
\eeq
with  eigenvalues
\beq
E_n=\left(1\pm \sqrt{(\delta_1-2 n)^2+\delta_2^2 n}\right)\epsilon_B\tr{\ \ for }n\geq1,
\eeq
as given in Eq. (5) in the main text. Besides the usual macroscopic LL degeneracy in the $xy$ plane for each LL, there is no additional degeneracy, including for the $n=0$ LL; this is in contrast to the two-fold degeneracy of the zero-energy LL state of the graphene-bilayer \cite{McCann06ap}.


\begin{thebibliography}{99}

\bibitem{Murakami07} S. Murakami, New J. Phys. \textbf{9}, 356 (2007).
\bibitem{Wan11} X. Wan, A. M. Turner, A. Vishwanath, and S. Y. Savrasov, Phys. Rev. B \textbf{83}, 205101 (2011).
\bibitem{Weng15a} H. Weng, C. Fang, Z. Fang, B. A. Bernevig, and X. Dai, Phys. Rev. X \textbf{5}, 011029 (2015).
\bibitem{Huang15} S.-M. Huang, S.-Y. Xu, I. Belopolski, C.-C. Lee, G. Chang, B. Wang, N. Alidoust, G. Bian, M. Neupane, C. Zhang, S. Jia, A. Bansil, H. Lin, and M. Z. Hasan, Nat. Commun. \textbf{6}, 8373 (2015).
\bibitem{Xu15} S.-Y. Xu, I. Belopolski, N. Alidoust, M. Neupane,  G. Bian, H. Shin-Ming, H. Zheng, J. Ma, D. S. Sanchex, B. Wang, A. Bansil, F. Chou, P. P. Shibayev, H. Lin, S. Jia, and M. Z. Hasan, Science \textbf{349}, 613 (2015).
\bibitem{Lu15} L. Lu, Z. Wang, D. Ye, L. Ran, L. Fu, J. D. Joannopoulos, and M. Solja\u{c}i\'{c}, Science \textbf{349}, 622 (2015).
\bibitem{Lv15a} B. Q. Lv, H. M. Weng, B. B. Fu, X. P. Wang, H. Miao, J. Ma, P. Richard, X. C. Huang, L. X. Zhao, G. F. Chen, Z. Fang, X. Dai, T. Qian, and H. Ding, Phys. Rev. X \textbf{5}, 031013 (2015).
\bibitem{Lv15b}  B. Q. Lv, N. Xu, H. M. Weng, J. Z. Ma, P. Richard, X. C. Huang, L. X. Zhao, G. F. Chen, C. E. Matt, F. Bisti, V. N. Strocov, J. Mesot, Z. Fang, X. Dai, T. Qian, M. Shi, and H. Ding,  Nat. Phys. \textbf{11}, 724 (2015).
\bibitem{Yang15} L. Yang, Z. Liu, Y. Sun, H. Peng, H. Yang, T. Zhang, B. Zhou, Y. Zhang, Y. Guo, M. Rahn, D. Prabhakaran, Z. Hussain, S. Mo, C. Felser, B. Yan, and Y. Chen,  Nat. Phys. \textbf{11}, 728 (2015).

\bibitem{Novo05} K. S. Novoselov, A. K. Geim, S. V. Morozov, D. Jiang, M. I. Katsnelson, I. V. Grigorieva, S. V. Dubonos, and A. A. Frisov, Nature (London) \textbf{438}, 197 (2005).
\bibitem{Gusynin05} V. P. Gusynin and S. G. Sharapov, Phys. Rev. Lett. \textbf{95}, 146801 (2005).
\bibitem{Zhang05} Y. Zhang, Y. W. Tan, H. L. Stormer, and P. Kim, Nature (London) \textbf{438}, 7065 (2005).

\bibitem{Volovik87} G. E. Volovik, JETP Lett. \textbf{46}, 98 (1987).
\bibitem{Volovik03} See, e.g., G. E. Volovik, \textit{The Universe in a Helium Droplet} (Oxford Science Publications, Oxford, 2003).
\bibitem{Haldane04} F. D. M. Haldane, Phys. Rev. Lett. \textbf{93}, 206602 (2004).
\bibitem{Klimkhamer05} F. R. Klinkhamer and G. E. Volovik, Int. J. Mod. Phys. A \textbf{20}, 2795 (2005).
\bibitem{Yang11} K.-Y. Yang, Y.-M. Lu, and Y. Ran, Phys. Rev. B \textbf{84}, 075129 (2011).
\bibitem{Burkov11a} A. A. Burkov and L. Balents, Phys. Rev. Lett. \textbf{107}, 127205 (2011).

\bibitem{Thouless82} D. J. Thouless, M. Kohmoto, M. P. Nightingale, and M. den Nijs, Phys. Rev. Lett. \textbf{49}, 405 (1982).
\bibitem{Hasan10} M.Z. Hasan and C.L. Kane, Rev. Mod. Phys. \textbf{82}, 3045 (2010).
\bibitem{Qi11} X.-L. Qi and S.-C. Zhang,  Rev. Mod. Phys. \textbf{83}, 1057 (2011).
\bibitem{Burkov11b}A. A. Burkov, M. D. Hook and L. Balents, Phys. Rev. B \textbf{84}, 235126 (2011).
\bibitem{Phillips14} M. Phillips and V. Aji, Phys. Rev. B \textbf{90}, 115111 (2014).
\bibitem{Weng15b} H. Weng Y. Liang, Q. Xu, R. Yu, Z. Fang, X. Dai, and Y. Kawazoe, Phys. Rev. B \textbf{92}, 045108 (2015).
\bibitem{Fang15} C. Fang, Y. Chen, H.-Y. Kee, and L. Fu, Phys. Rev. B \textbf{92}, 081201(R) (2015).
\bibitem{Mullen15} K. Mullen, B. Uchoa, and D. T. Glatzhofer, Phys. Rev. Lett. \textbf{115}, 026403 (2015).
\bibitem{Kim15} Y. Kim, B. J. Wieder, C. L. Kane, and A. M. Rappe, Phys. Rev. Lett. \textbf{115}, 036806 (2015).
\bibitem{Yu15} R. Yu, H. Weng, Z. Fang, X. Dai, and X. Hu, Phys. Rev. Lett. \textbf{115}, 036807 (2015).
\bibitem{Chen15} Y. Chen, Y. Xie, S. A. Yang, H. Pan, F. Zhang, M. L. Cohen, S. Zhang, Nano Lett. \textbf{15} (10), 6974 (2015).
\bibitem{Rhim15} J.-W. Rhim and Y. B. Kim, Phys. Rev. B \textbf{92}, 045126 (2015).
\bibitem{Heikkila15} T. E. Heikkila and G. E. Volovik, New J. Phys. \textbf{17}, 093019 (2015).
\bibitem{Xie15} L. S. Xie, L. M. Schoop, E. M. Seibel, Q. D. Gibson, W. Xie, and R. J. Cava, APL Mater. \textbf{3}, 083602 (2015).
\bibitem{Chan15}Y.-H. Chan, C.-K. Chiu, M.Y. Chou, and A. P. Schnyder, Phys. Rev. B \textbf{93}, 205132 (2016).
\bibitem{Bian15} G. Bian, T.-R. Chang, H. Zheng, S. Velury, S.-Y. Xu, T. Neupert, C.-K. Chiu, S.-M. Huang, D. S. Sanchez, I. Belopolski, N. Alidoust, P.-J. Chen, G. Chang, A. Bansil, H.-T. Jeng, H. Lin, and M. Z. Hasan, Phys. Rev. B \textbf{93}, 121113(R) (2016).
\bibitem{Bian16} G. Bian, T.-R. Chang, R. Sankar, S.-Y. Xu, H. Zheng, T. Neupert, C.-K. Chiu, S.-M. Huang, G. Chang, I. Belopolski, D. S. Sanchez, M. Neupane, N. Alidoust, C. Liu, B. Wang, H.-T. Jeng, A. Bansil, F. Chou, H. Lin, and M. Z. Hasan, Nat. Commun. \textbf{7}, 10556 (2016).
\bibitem{Yamakage16}A. Yamakage, Y. Yamakawa, Y. Tanaka, and Y. Okamoto, J. Phys. Soc. Jpn. \textbf{85}, 013708 (2016).
\bibitem{Ezawa16} M. Ezawa, Phys. Rev. Lett. \text{116}, 127202 (2016).
\bibitem{Wang16} J.-T. Wang, H. Weng, S. Nie, Z. Fang, Y. Kawazoe, and C. Chen, Phys. Rev. Lett. \textbf{116}, 195501 (2016).
\bibitem{Bzdusek16} T. Bzdu\v{s}ek, Q. Wu, A. R\"{u}egg, M. Sigrist, and A. A. Soluyanov, Nature (London) \textbf{538}, 75 (2016).
\bibitem{Yan16} Z. Yan and Z. Wang, Phys. Rev. Lett. \textbf{117}, 087402 (2016).
\bibitem{Chan16} C.-K. Chan, Y.-T. Oh, J. H. Han, and P. A. Lee, Phys. Rev. B \textbf{94}, 121106 (2016).





\bibitem{Saffman92} P. G. Saffman, \textit{Vortex Dynamics} (Cambridge University Press, Cambridge, England, 1992).
\bibitem{Cooper99} N. R. Cooper, Phys. Rev. Lett. \textbf{82}, 1554 (1999).

\bibitem{McCann06} E. McCann and V. I. Fal’ko, Phys. Rev. Lett. \textbf{96}, 086805 (2006).
\bibitem{Gail11} R. de Gail, M. O. Goerbig, F. Guinea, G. Montambaux, and A. H. Castro Neto, Phys. Rev. B \textbf{84}, 045436 (2011).

\bibitem{Nielsen81} H. B. Nielsen and M. Ninomiya, Nucl. Phys. \textbf{B185}, 20 (1981).

\bibitem{Mon09} G. Montambaux, F. Piechon, J.-N. Fuchs, and M. O. Goerbig, Eur. Phys. J. B \textbf{72}, 509 (2009).




\bibitem{supp} See Supplemental Material at [url] for a derivation of general vortex ring Hamiltonians, definitions of topological indicesm details of the tight-binding model and detailed solution to the Landau level spectrum, which includes Ref. [49].
\bibitem{Lim15} L.-K. Lim, J.-N. Fuchs, and G. Montambaux, Phys. Rev. A \textbf{92}, 063627 (2015).



\bibitem{foot1} For spin-split bands and Pauli matrices acting only on the sublattice degree of freedom, the effective Hamiltonian under time-reversal transformation, $H(\mathbf{k})\rightarrow  H^*(-\mathbf{k})$, and under inversion transformation, $H(\mathbf{k})\rightarrow \sigma_x H(-\mathbf{k}) \sigma_x$; neither of which is a symmetry of the Hamiltonian. 

\bibitem{Hosur12} P. Hosur, Phys. Rev. B \textbf{86}, 195102 (2012).
\bibitem{Haldane14} F. D. M. Haldane, arXiv:1401.0529.
\bibitem{foot2} A microscopic derivation of the chiral surface states can be constructed in analogy to Ref. \cite{Hosur12}.

\bibitem{foot3} In Figs. 4b-c, the putative surface states on the surface Brillouin zone \textit{parallel} to the nodal ring plane are not chiral surface states that cross the bulk energy gap. Their existence is due to boundary effects. In this case, only high energy states which deviate sufficiently from the bulk energy levels are localized on the surface. 

\bibitem{GM15} D. Gosalbez-Martinez, I. Souza, and D. Vanderbilt, Phys. Rev. B \textbf{92}, 085138 (2015).

\bibitem{Parameswaran14} S. A. Parameswaran, T. Grover, D. A. Abanin, D. A. Pesin, and A. Vishwanath, Phys. Rev. X \textbf{4}, 031035 (2014).
\bibitem{Gorbachev14} R. V. Gorbachev, J. C. W. Song, G. L. Yu, A. V. Kretinin, F. Withers, Y. Cao, A. Mishchenko, I. V. Grigorieva, K. S. Novoselov, L. S. Levitov, and A. K. Geim, Science \textbf{346}, 448 (2014).

\bibitem{Konig07} M. K\"{o}nig, S. Wiedmann, C. Br\"{u}ne, A. Roth, H. Buhmann, L. W. Molenkamp, Q.-X. Liang, and S.-C. Zhang, Science \textbf{318}, 766 (2007).

\bibitem{Fuchs10} J.-N. Fuchs, F. Piechon, M. O. Goerbig, and G. Montambaux, Eur. Phys. J. B \textbf{77}, 351 (2010).

\bibitem{Weng16s} See, e.g., H. Weng, X. Dai, and Z. Fang, J. Phys. Condens. Matter \textbf{28}, 303001 (2016).
\bibitem{Dub15} T. Dub\u{c}ek, C. J. Kennedy, L. Lu, W. Ketterle, M. Solja\u{c}i\'{c}, and H. Buljan, Phys. Rev. Lett. \textbf{114}, 225301 (2015).
\bibitem{Xu16} Y. Xu and C. Zhang, Phys. Rev. A \textbf{93}, 063606 (2016).
\bibitem{DZhang16} D.-W. Zhang, Y. X. Zhao, R.-B. Liu, Z.-Y. Xue, S.-L. Zhu, and Z. D. Wang, Phys. Rev. A \textbf{93}, 043617 (2016).

\bibitem{Dora13} B. D\'{o}ra, I. F. Herbut, R. Moessner, Phys. Rev. B \textbf{88}, 075126 (2013).
\bibitem{Yang14} B.-J. Yang, E.-G. Moon, H. Isobe, and N. Nagaosa, Nat. Phys. \textbf{10}, 774 (2014).

\bibitem{Roy16}B. Roy, arXiv:1607.07867.
\bibitem{Sur16} S. Sur and R. Nandkishore, arXiv:1608.08198.

\end{thebibliography}

\begin{thebibliography}{99}
\bibitem{Burkov11ap} A. A. Burkov and L. Balents, Phys. Rev. Lett. \textbf{107}, 127205 (2011).
\bibitem{Lim15} L.-K. Lim, J.-N. Fuchs, and G. Montambaux, Phys. Rev. A \textbf{92}, 063627 (2015).
\bibitem{McCann06ap} E. McCann and V. I. Fal’ko, Phys. Rev. Lett. \textbf{96}, 086805 (2006).
\end{thebibliography}
\end{document}